\documentclass{article}
\usepackage{amssymb,amsmath,amsthm,amsbsy,xcolor}
\usepackage{epsfig,amssymb,amsmath,textcomp,caption,verbatim,gensymb,wrapfig,mathtools, tabularx}
\usepackage{epstopdf, ltablex,booktabs,layouts}
\usepackage{graphicx}
\usepackage{amsfonts,braket}
\usepackage[font={small}]{caption}
\usepackage{subfig}
\usepackage[colorlinks=true,citecolor=blue]{hyperref}
\usepackage{cleveref}
\usepackage[square,comma,numbers,compress]{natbib}

\begin{document}
	
	\title{Quantum Soliton Scattering Manifolds}
	\author{Chris Halcrow\thanks{email: c.j.halcrow(at)leeds.ac.uk} \\ \\ {\it School of Mathematics,} \\ 
		{\it University of Leeds,}\\
		{\it Leeds, LS2 9JT, UK. } }

\date{\today}

\maketitle

{\bf Abstract}
We consider the quantum multisoliton scattering problem. For BPS theories one truncates the full field theory to the moduli space, a finite dimensional manifold of energy minimising field configurations, and studies the quantum mechanical problem on this. Non-BPS theories -- the generic case -- have no such obvious truncation. We define a quantum soliton scattering manifold as a configuration space which satisfies asymptotic completeness and respects the underlying classical dynamics of slow moving solitons. Having done this, we present a new method to construct such manifolds. In the BPS case the dimension of the $n$-soliton moduli space $\mathcal{M}_n$ is $n$ multiplied by the dimension of $\mathcal{M}_1$. We show that this scaling is not necessarily valid for scattering manifolds in non-BPS theories, and argue that it is false for the Skyrme and baby-Skyrme models. In these models, we show that a relative phase difference can generate a relative size difference during a soliton collision. Asymptotically, these are zero and non-zero modes respectively and this new mechanism softens the dichotomy between such modes. Using this discovery, we then show that all previous truncations of the 2-Skyrmion configuration space are unsuitable for the quantum scattering problem as they have the wrong dimension. This gives credence to recent numerical work which suggests that the low-energy configuration space is 14-dimensional (rather than 12-dimensional, as previously thought). We suggest some ways to construct a suitable manifold for the 2-Skyrmion problem, and discuss applications of our new definition and construction for general soliton theories. 
	
\section{Introduction}
	
Solitons appear in a wide range of physical theories: as nuclei in the Skyrme model \cite{Sk61}, vortices in superconductors \cite{GL50}, monopoles in GUTS \cite{tH74} and spin whirls in magnetic materials \cite{Yu10,BY89}, to name a few examples. The moduli space approximation, developed to describe low energy soliton motion, is even used to model extremal black hole dynamics \cite{GR86}. In this approximation the full field theory is truncated and only the most important degrees of freedom are kept \cite{Man82}. For BPS theories (those with no classical forces between solitons) there is an obvious truncation - one only permits configurations which satisfy the first order BPS equations, which are in turn equivalent to the static second order Euler-Lagrange equations. These configurations, which all have the same energy, form a manifold called the moduli space. The moduli space of $n$ solitons is denoted $\mathcal{M}_n$ and the dynamics of solitons is then described by free motion on $\mathcal{M}_n$.

For certain theories (such as almost-critically coupled vortices \cite{DS94}) the validity of the moduli space approximation can be rigorously proved at small soliton velocities. It also makes sense physically: there is an effective potential on the space of all configurations; configurations in the moduli space parametrize the bottom of this potential and, since total energy is conserved and kinetic energy is small, motion must take place near here.

In the most widely studied systems, the single soliton moduli space $\mathcal{M}_1$ contains the 1-soliton and its orbit under the symmetry group of the underlying theory. For example, for the $SU(2)$-monopole system, the moduli space  $\mathcal{M}_1$ is isomorphic to $\mathbb{R}^3 \times U(1)$; $\mathbb{R}^3$ arises from translations in physical space and $U(1)$ from transformations in target space. Generically, the $n+m$-soliton moduli space has regions where the soliton breaks up into an $n$-soliton and an $m$-soliton so that, in this region, $\mathcal{M}_{n+m}$ is approximately given by the product of $\mathcal{M}_n$ and $\mathcal{M}_m$. We'll call this idealized part of the manifold the asymptotic submanifold. When the solitons are closer together, the structure of $\mathcal{M}_{n+m}$ deforms significantly, but the asymptotic picture can be helpful. For instance, this picture suggests that
\begin{align} \label{dim}
&\text{Dim}\left(\mathcal{M}_{n+m}\right) =  \text{Dim}\left(\mathcal{M}_{n}\right) + \text{Dim}\left(\mathcal{M}_{m}\right) \nonumber \\
\text{and hence } &\text{Dim}\left(\mathcal{M}_{n}\right) = n \text{Dim}\left(\mathcal{M}_{1}\right) 
\end{align}
and this is true for a wide range of theories. In BPS theories, this relation can often be proved rigorously using index theorem calculations. 

For many applications, one wishes to solve the quantum multisoliton scattering problem. For BPS theories, there is a metric on the moduli space $\mathcal{M}_{n+m}$ induced by the field theory, giving a Schr\"odinger equation
\begin{equation} \label{BPSschro}
i\hbar \frac{\partial \Psi}{\partial t} = -\frac{\hbar^2}{2} \Delta \Psi \, ,
\end{equation}
where $\Delta$ is the Laplace-Beltrami operator on $\mathcal{M}_{n+m}$. To construct the ingoing states,  we look to the asymptotic submanifolds. On them we can construct one-particle states, the total wavefunction equal to the product of free wavefunctions on $\mathcal{M}_n$ and $\mathcal{M}_m$. One then solves \eqref{BPSschro}. Although conceptually simple, the calculation has only been undertaken in a few cases including $SU(2)$-monopoles \cite{Sch91} and critically coupled vortices \cite{Sam92}.

In contrast, the theory of non-BPS soliton scattering is barely developed. Although there is a moduli space for each topological sector, they usually do not allow the solitons to separate into their individual identities and are thus too small to describe scattering problems. Instead, one attempts to write down an approximate configuration space of static low energy configurations which we'll denote as $\mathcal{S}_{n,m}$. This notation highlights that the appropriate space depends on the topological degree of each particle. Configurations on $\mathcal{S}_{n,m}$ may have different energies and so the Schr\"odinger equation becomes
\begin{equation*} 
i\hbar \frac{\partial \Psi}{\partial t} = -\frac{\hbar^2}{2} \Delta \Psi  + V\Psi\, ,
\end{equation*}
where $V$ is the static energy of the configurations on $\mathcal{S}_{n,m}$. Configurations in $\mathcal{S}_{n,m}$ do not satisfy a mathematical condition, instead one must rely on physical intuition to construct the manifold. There is no consensus on what the manifold should look like. For example, several such manifolds have been considered for $1+1$ scattering in the Skyrme model. These have involved the product approximation \cite{JJ85}, instanton approximation \cite{AM93} and modeling the space as a union of gradient flows from the unstable $B=2$ hedgehog \cite{Man88}. All of these rely on the intuition of \eqref{dim}, since a single Skyrmion has 6 moduli, $\mathcal{S}_{1,1}$ should be 12-dimensional. Later in this paper, we will argue that this is false: the quantum Skyrmion-Skyrmion scattering manifold has dimension greater than 12. Hence the previous models are unsuitable to describe nucleon-nucleon scattering in the Skyrme model. Key to this argument is asymptotic completeness, a condition which ensures that ingoing particles end up becoming outgoing particles. We will study this condition for soliton models later on.

In this paper, we will carefully consider non-BPS quantum soliton scattering. We first propose a definition of a quantum soliton scattering manifold. The definition contains the minimal requirements for the manifold to support multi-soliton scattering. This in hand, we develop a new construction for $\mathcal{S}_{1,1}$, based on scattering paths generated using the asymptotic submanifold as initial data. Although this construction appears almost tautological - using scattering paths to generate a manifold which is then used to approximate scattering paths - it can provide vital insights. In particular, we apply the new construction to baby Skyrmions, where we show the naive manifold must be enlarged to include relative size fluctuations. Having done this, we apply the same insights to Skyrmions in 3-dimensions and show that a quantum Skyrmion-Skyrmion scattering manifold must have dimension greater than 12. We conclude by speculating on the true structure of $\mathcal{S}_{n,m}$ for Skyrmions and uses for our construction in other settings.

\section{Quantum soliton scattering manifolds}

Before considering the quantum scattering problem for solitons, we will consider the quantum scattering problem for particles. Roughly, the problem is to solve the Schr\"odinger equation
\begin{equation} \label{QMsch}
i\frac{\partial \Psi}{\partial t} = \mathcal{H}\Psi \equiv (\mathcal{H}_0 + \mathcal{H}_I)\Psi \, ,
\end{equation}
where $\Psi$ is a wavefunction and $\mathcal{H}$ is the Hamiltonian operator which can be split into free and interacting parts. Quantum time evolution can be formally read off from \eqref{QMsch}. It is
\begin{equation*}
\Psi(t) = e^{-i \mathcal{H} t} \Psi(0) \, .
\end{equation*}
For a scattering problem, one of the initial (or ingoing) conditions is that the wavefunction is an eigenfunction of the free Hamiltonian at early time. That is
\begin{equation} \label{ingoing}
\lim_{t\to-\infty} \Psi = \Psi_0^- , \qquad \text{where} \quad \mathcal{H}_0\Psi_0^- = k^2 \Psi_0^-
\end{equation}
for some eigenvalue $k^2$. Physically these states represent particles far from each other, not interacting. Each particle in the system can be described by a one-particle state and $\Psi_0^-$ is a product of these states. From early in the study of quantum mechanics, people realized the importance of a condition called asymptotic completeness. It says that if the early-time state $\Psi_0^-$ belongs to the Hilbert space of $\mathcal{H}_0$, so should its late time state. That is $\Psi_0^+ = \lim_{t\to\infty} e^{-i \mathcal{H} t}\Psi_0^-$ should also be an eigenstate of $\mathcal{H}_0$. Another way of saying this: any free ingoing state gets mapped to a free outgoing state under time evolution. This condition was proved rigorously for many particle-potential scattering problems by Ikebe \cite{Ike60}, then generalized for three and $n$ particles by Fadeev \cite{Fad65} and Hepp \cite{Hep69} respectively.

We'll now focus on topological solitons. We study field theories with an integer-valued topological charge which splits the configuration space into disjoint subspaces, labeled by the charge. We denote the total configuration space of the field theory  $\mathcal{C}^\infty$, its disjoint subspaces $\mathcal{C}_n^\infty$ and use $\mathcal{C}$ when discussing regions of this infinite-dimensional space. The manifold of minimal energy solutions with topological charge $n$ is called the moduli space and is denoted $\mathcal{M}_n$. We will go on to define and discuss scattering spaces. We use $\mathcal{S}$ to denote the scattering spaces, to maintain a distinction between these and moduli spaces.

Our definitions will rely on the notion of ingoing and outgoing regions, and we denote these as $\mathcal{C}^\text{in} \subset \mathcal{C}^\infty$ and $\mathcal{C}^\text{out}\subset \mathcal{C}^\infty$. For particle systems, these are regions when the particles are widely separated. For some soliton systems these regions are straightforward to define. For instance, a critically coupled vortex has a position given by the zeros of the Higgs field. One can track the positions of two vortices by tracking the positions of these zeroes on $\mathbb{R}^2$. If the vortices are positioned in the first and third quadrants of the complex plane and sent towards one another with momentum along the $x$-axis, the outgoing vortices take their positions in the second and fourth quadrants. Since the solitons have position, we can define a relative separation. This separation should be large in the ingoing and outgoing regions.  Hence, in the vortex example, one can say that a configuration belongs to $\mathcal{C}^\text{in}$ if the zeroes of the Higgs field are located in the first and third quadrants and the distance between the zeroes is larger than some fixed separation. A configuration belongs to $\mathcal{C}^\text{out}$ if the zeroes are located in the second and third quadrants, and their difference is greater than some fixed separation. In other examples, the ingoing and outgoing spaces will be equal.

With a notion of ingoing and outgoing regions, we can define an asymptotic submanifold $\mathcal{M}^\text{asy}_{n,m}$. Consider the space of configurations with topological charge $n+m$. The asymptotic submanifold of an $(n+m)$-soliton system is a subspace of $\mathcal{C}^\infty_{n+m}$ where the solitons are widely separated and look like two individual solitons. The asymptotic submanifold is then approximately isomorphic to the product of the moduli spaces $\mathcal{M}_n$ and $\mathcal{M}_m$,
\begin{equation*}
\mathcal{M}^\text{asy}_{n,m} \cong \mathcal{M}_n \times \mathcal{M}_{m} \, .
\end{equation*}
This expression is only valid when the solitons are widely separated. If the underlying theory has a symmetry group $G$, we can rewrite this as
\begin{equation*}
\mathcal{M}^\text{asy}_{n,m} \cong G \times \widetilde{\mathcal{M}}_{n,m}^\text{asy} \, ,
\end{equation*}
and we call $\widetilde{\mathcal{M}}_{n,m}$ the reduced asymptotic manifold.

Our initial ingoing state, analogous to \eqref{ingoing} is a product of free quantum solitons. There is a good understanding of what a quantum n-soliton is -- a free one-particle state on $\mathcal{M}_n$. Hence, a free two-soliton state is defined on the product $\mathcal{M}_n \times \mathcal{M}_m$, as the product of the one-particle states. The ingoing particles in our scattering theory are the quantum states on the ingoing asymptotic submanifold. Our definition of a quantum soliton scattering manifold demands that we have asymptotic completeness. That is, every ingoing state is linked to an outgoing state by time-evolution where the outgoing states are also defined on an asymptotic submanifold. Assuming that we can define an ingoing and outgoing region, we are now ready for the definition.

Consider a $G$-principle bundle
\begin{equation*}
G \longrightarrow \mathcal{S}_{n,m} \xrightarrow{\, \ \pi \ } \widetilde{\mathcal{S}}_{n,m} \, ,
\end{equation*}
\begin{flushleft}
where $\widetilde{\mathcal{S}}_{n,m}$ is a submanifold of configuration space, called the reduced scattering manifold, and $G$ is the symmetry group of the underlying field theory. Then $\mathcal{S}_{n,m}$ is a \emph{quantum soliton scattering manifold} (QSS$\mathcal{M}$) for describing $n$-soliton, $m$-soliton scattering if: \linebreak
$a)$ Curves in $\mathcal{S}_{n,m}$ approximate low energy classical dynamics of $(m+n)$-solitons. \linebreak
$b)$ The ingoing region of $\mathcal{S}_{n,m}$ contains the asymptotic submanifold $\mathcal{M}_n\times\mathcal{M}_m$. This is linked, via paths in $\mathcal{S}_{n,m}$, to the outgoing region of $\mathcal{S}_{n,m}$. This region should also contain a copy of $\mathcal{M}_n\times\mathcal{M}_m$.
\end{flushleft}

 Condition $a)$ simply states that, since the entire quantum soliton approximation relies on a manifold of classical configurations, the manifold should respect the classical picture. Time evolution is taken from the equations of motion derived from the Lagrangian. If our initial data is $\phi_0 \in \mathcal{C}$ and $\dot{\phi}_0 \in T_{\phi_0} \mathcal{C}$ (the tangent space of configuration space), then let $\mathcal{T}(\phi_0, \dot{\phi}_0 , t)$ be the solution of the equations of motion for that initial data, at time $t$. The one-dimension set of configurations
 \begin{equation*}
 \gamma_{\phi_0,\dot{\phi_0}} =  \left\{ \mathcal{T}\left(\phi_0,\dot{\phi}_0,t\right) : t \in \mathbb{R} \right\}\, , 
 \end{equation*}
 is a scattering path. Condition $a)$ then states that these paths are well approximated by paths in $\mathcal{S}_{n,m}$.

 Condition $b)$ is required to satisfy asymptotic completeness. If the outgoing asymptotic manifold does not contain this factor, we cannot form outgoing one-particle states and so we cannot answer questions about the quantum soliton scattering. For instance, we cannot calculate the S-matrix for $m+n \to m+n$ scattering. For well understood BPS theories, the moduli space is a quantum soliton scattering manifold.

\begin{figure}[h!]
	\begin{center}
		\includegraphics[width=0.95\textwidth]{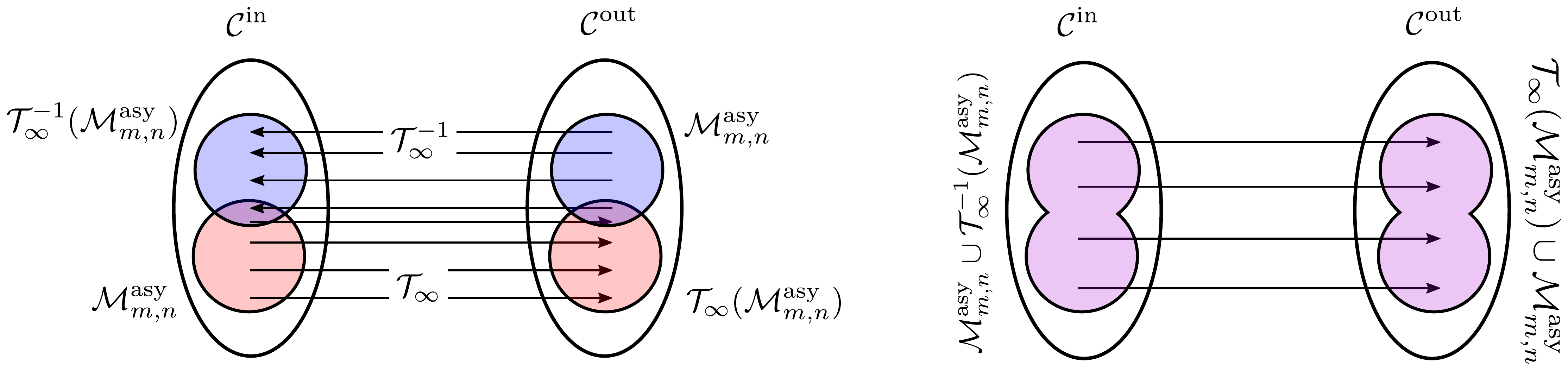} 
	\end{center}
	\caption{A sketch of the construction of a QSS$\mathcal{M}$, using time evolution and the asymptotic submanifold. In the righthand figure, the source and image contain a copy of $\mathcal{M}^\text{asy}_{m,n}$.  } \label{construction}
\end{figure}

Note that the definition of a QSS$\mathcal{M}$ immediately reveals some of the structure of $\mathcal{S}_{n,m}$. It is required to contain a copy of $\mathcal{M}^\text{asy}_{m,n}$ in the ingoing region, and must also contain scattering paths. Hence $\mathcal{M}_{n,m}$ should contain all paths whose initial data is contained in $\mathcal{M}^\text{asy}_{m,n}$. That is, the set
\begin{equation} \label{gamma}
\Gamma_{\mathcal{M}^\text{asy}_{m,n}} = \left\{ \gamma_{\phi_0,\dot{\phi_0}} :\phi_0 \in \mathcal{M}^\text{asy}_{m,n}, \, \, \dot{\phi}_0 \in T_{\phi_0}\mathcal{M}^\text{asy}_{m,n}  \right\} \, .
\end{equation}
In fact, this gives an initial guess for the entirety of $\mathcal{S}_{m,n}$. If this contains an outgoing copy of $\mathcal{M}^\text{asy}_{m,n}$, it is a QSS$\mathcal{M}$. This question is easy to resolve by investigating the paths with initial data in the asymptotic submanifold. However, if the paths do not contain $\mathcal{M}^\text{asy}_{m,n}$, they still might help us construct a QSS$\mathcal{M}$. Suppose that each path in \eqref{gamma} links an ingoing and outgoing configuration. Hence there is a set $\mathcal{T}_\infty(\mathcal{M}^\text{asy}_{m,n}) \subset \mathcal{C}^\text{out}$ which is linked to $\mathcal{M}^\text{asy}_{m,n} \subset \mathcal{C}^\text{in}$. Similarly, there is also an outgoing set $\mathcal{M}^\text{asy}_{m,n} \subset \mathcal{C}^\text{out}$ which is linked to some set $\mathcal{T}^{-1}_\infty(\mathcal{M}^\text{asy}_{m,n}) \subset \mathcal{C}^\text{in}$. To find $\mathcal{T}^{-1}_\infty(\mathcal{M}^\text{asy}_{m,n})$ you simply consider time evolution of all configurations in the outgoing asymptotic manifold. The manifolds $\mathcal{M}^\text{asy}_{m,n}$ and $\mathcal{T}^{-1}(\mathcal{M}^\text{asy}_{m,n})$ locally (at each separation) look like $V_1 \times G$ and $V_2 \times G$, where $V_1$ and $V_2$ are vector spaces. The sum of these, $(V_1 + V_2) \times G$ provides the base of an ingoing manifold $\mathcal{S}^\text{in}$. This contains $\mathcal{M}^\text{asy}_{m,n}$ and is linked, by time evolution, to another manifold $\mathcal{S}^\text{out}$ which also contains $\mathcal{M}^\text{asy}_{m,n}$. The union of all paths joining $\mathcal{S}^\text{in}$ and $\mathcal{S}^\text{out}$ is then a QSS$\mathcal{M}$. This construction is shown pictorially in Figure \ref{construction}.

Unfortunately, this construction is indirect. It would be difficult to construct an explicit configuration space with a metric and potential (needed for the quantum calculation), except in simple cases such as kinks in one-dimension. Instead, the hope is that one can use this construction to discover properties of a QSS$\mathcal{M}$, such as its dimension, and use this information as inspiration to generate an approximate configuration space. If this is achieved, one can then study the quantum mechanical problem. 

The construction is semi-classical since we build our manifold using classical scattering paths, and is valid at small soliton velocities. We use these paths to find a metric and potential which is then used in a quantum mechanical problem. Note that this is different than \emph{the} semi-classical method used in, e.g., \cite{Bra87}. There, the scattering amplitude is written in terms of classical quantities such as time-delays. Although these techniques are useful in one-dimension and scale well in higher dimensions, the expansion is only valid at high velocities. In this region, the scattering amplitude tends to the classical result. As such, the textbook semi-classical method is not useful for slow moving solitons in more than one-dimension. Our construction, which relies only on an understanding of the classical time evolution of a finite set of configurations can be used instead.

Our description is rather informal.  This is partially because solitons are hard to discuss in general. Generating the asymptotic manifold is trivial for some systems and intricate for others. The time-evolution operator $\mathcal{T}$ may be ill-defined for some systems, or produce very different manifolds depending on initial momenta. We have also assumed that the configuration space is well behaved. Singularities, such as those found in \cite{Eto06}, may cause difficulties. In the upcoming Sections, we will focus on three systems and on $1+1$ scattering. This allows us to be far more explicit.

\section{The sine-Gordon model}
	
To start exploring the new definition and construction let us consider a well understood system: two sine-Gordon kinks, where the questions we'll ask have well known answers. The sine-Gordon model has Lagrangian
\begin{equation*}
\mathcal{L} =\tfrac{1}{2} \dot \phi(x,t)^2 - \tfrac{1}{2}\phi_{x}^2(x,t) - (1-\cos(\phi(x,t))) \, .
\end{equation*}
This has Lorentzian symmetry, and so we expect $E_1 = \mathbb{R}$ to be the fiber of the scattering manifold. The Euler-Lagrange equations are
\begin{equation} \label{sgeom}
-\ddot \phi + \phi_{xx} = \sin(\phi) \, ,
\end{equation}
and the single kink solutions satisfying the static version of \eqref{sgeom} is given by
\begin{equation*}
\phi(x,t;X) = 4 \tan^{-1}\left( e^{x-X} \right)  \iff \tan(\phi(x,t;X)/4) = e^{x-X} \, .
\end{equation*}
The kink has a single moduli $X$ interpreted as its position, where $\phi = \pi$. The moduli space of a single kink is $\mathcal{M}_1 \cong \mathbb{R}$ and so the asymptotic submanifold of two kinks is given by
\begin{equation*}
\mathcal{M}_{1,1}^\text{asy} \cong \frac{\mathbb{R}\times\mathbb{R}}{\mathbb{Z}_2}  \cong \mathbb{R} \times \mathbb{R}^+ = \mathbb{R} \times \widetilde{\mathcal{M}}^\text{asy}_{1,1} \, .
\end{equation*}
Here, $\mathbb{R}$ represents the center-of-mass of the two kinks,  while $\mathbb{R}^+$ represents separation. The ingoing region of configuration space is where the kinks are separated by at least some constant $R$. The reduced asymptotic submanifold $\widetilde{M}_{1,1}$ is the interval $[R,\infty)$, which represents the separation of the kinks.

 There are no static 2-kink solutions but
\begin{equation*}
\tan\left(\phi_2(x;X_1,X_2)/4\right) = e^{x-X_1} - e^{-x+X_2} \, .
\end{equation*}
is an approximate solution when $|X_1 - X_2|$ is large. Defining the center of mass as $C = \tfrac{1}{2}(X_1 + X_2)$ and the relative position as $2X = (X_1 - X_2)$, the 2-kink solution becomes
\begin{equation*} \label{sol}
\tan\left(\phi_2(x;C,X)/4\right) =  e^{(x-C)-X} - e^{-(x-C)-X} \, .
\end{equation*}

\begin{figure}[h!]
	\begin{center}
		\includegraphics[width=0.9\textwidth]{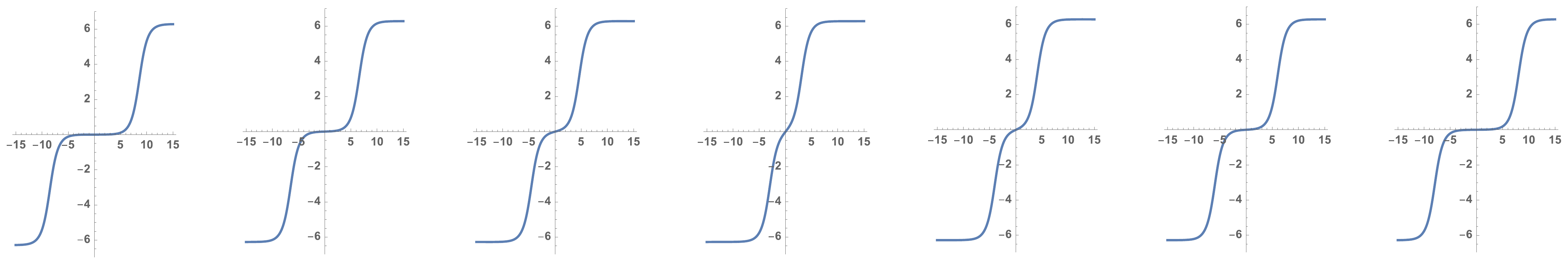} 
	\end{center}
	\caption{ Kink-kink scattering in the sine-Gordon model. The initial configuration has $X=8$ and $v=0.1$. The time evolution, read left to right, takes place over 140 time units. The two individual kinks can be clearly seen initially. They deform while colliding before separating and regaining their individual identities once again.  } \label{fig:Kinks}
\end{figure}

We wish to construct a configuration space using the scattering paths whose initial configurations are in the reduced asymptotic submanifold. But it is not guaranteed a priori that this will be a QSS$\mathcal{M}$. To check, we must investigate the scattering paths and ensure that the set of outgoing configurations contains $\widetilde{\mathcal{M}}_{1,1}^\text{asy}$. In this case, we must scatter two well separated kinks and check if two well separated kinks emerge after their collision. We do this  by taking $X = X_0 - v t$ and $C=0$ initially, giving the initial conditions
\begin{align*}
\tan( \phi_2(x,0) / 4)&=  \left( e^{x-X_0} - e^{-x-X_0} \right) \\ 
\dot \phi_2(x,0) \sec^2(\phi / 4) / 4 &=  v \left( e^{x-X_0} - e^{-x-X_\text{in}} \right)
\end{align*}
This says that at early time, the two kinks are separated by $2X_0$ and are moving towards one another, each with velocity $v$. We can solve these equations numerically and a solution is shown in Figure \ref{fig:Kinks}. The scattering reveals two things. First, the outgoing region is the same as the ingoing region, but where the kink momenta point away from the center of mass. Second, that the outgoing configurations do look like the configurations in the outgoing asymptotic submanifold. This fact could be quantified by equipping configuration space with an inner product, and comparing the numerical solutions with the known solutions from the asymptotic submanifold. Overall, the configurations produced in the scattering, fibered by $\mathbb{R}$, do parametrize a two-dimensional reduced QSS$\mathcal{M}$ as expected. The bundle has a global product structure so is trivial.

The numerical solution approximates the exact solution, which is known to be \cite{book}
\begin{align*} 
\phi_2(x,t) &= 4 \tan^{-1} \left( e^{\gamma(x-X(t))} - e^{-\gamma(x+X(t))} \right) \, , \\ \text{where } \, X(t) &= \gamma^{-1} \log\left(2 v^{-1} \cosh(v \gamma t) \right) 
\end{align*}
and $\gamma$ is the Lorentz factor $\gamma = \sqrt{1-v^2}$.

\section{The Baby Skyrme model}

The baby Skyrme model was proposed as a 2-dimensional analog of the full Skyrme model. To emphasize the similarity we write the Lagrangian as a nonlinear sigma model,
\begin{equation*} \label{Lag}
\mathcal{L} = \tfrac{1}{2} \partial_\mu\boldsymbol{\pi} \cdot \partial^\mu \boldsymbol{\pi} - \tfrac{1}{4}\left( \left(\partial_\mu \boldsymbol{\pi} \cdot \partial^\mu \boldsymbol{\pi}\right)^2 - \left(\partial_\mu \boldsymbol{\pi} \cdot \partial^\nu \boldsymbol{\pi} \right)\left(\partial_\nu \boldsymbol{\pi} \cdot \partial^\mu \boldsymbol{\pi}\right) \right) - m^2(1-\pi_3) \, ,
\end{equation*}
where $m$ is analogous to the dimensionless pion mass and $\boldsymbol{\pi}$ is a field satisfying $\boldsymbol{\pi} \cdot \boldsymbol{\pi} = 1$ so that $\boldsymbol{\pi} : \mathbb{R}^2 \to S^2$. It is also helpful to write $\mathcal{L}$ in terms of a single complex scalar field
\begin{equation*}
W = \frac{\pi_1 + i \pi_2}{1+ \pi_3} \, .
\end{equation*}
Then the Lagrangian takes the form
\begin{equation} \label{LagW}
\mathcal{L} = \frac{ \partial_\mu W \partial^\mu \overline{W} }{\left(1+ |W|^2 \right)^2} + \frac{\epsilon_{\alpha \beta \gamma } \epsilon^{\alpha \mu \nu} \partial^\beta \overline{W} \partial^\gamma W \partial_\mu \overline{W} \partial_\nu W }{\left(1+|W|^2\right)^4} - m^2\left( \frac{1-|W|^2}{1+|W|^2} \right) \, .
\end{equation}
The Lagrangian enjoys Lorentz symmetry and isorotational symmetry ($W \to e^{ia}W, a \in \mathbb{R}$). So we expect the scattering manifold to have the fiber $G = E_2 \times S^1$ where $E_2$ is the Euclidean group of the plane. For finite energy configurations the potential in \eqref{LagW} forces $W$ to tend to a constant value at infinity. This causes a one-point compactification of space and hence finite energy configurations are maps between 2-spheres. Maps of this kind have a conserved topological charge, $B$. A minimal energy configuration with charge $B$ is known as a $B$ baby Skyrmion. Time evolution arises from the Euler-Lagrange equations of  \eqref{LagW}. The explicit equations are unpleasant and are written out fully in \cite{PPZ92}. 

We will now construct a quantum soliton scattering manifold (QSS$\mathcal{M}$) for $1+1$ baby Skyrmion scattering. To do so, we first study the $B=1$ baby Skyrmion, and use it to construct the asymptotic submanifold $\mathcal{M}_{1,1}^\text{asy}$. This schematically looks like two widely separated baby Skyrmions. We will then study scattering paths whose initial data lies in $\mathcal{M}_{1,1}^\text{asy}$. Using this, and the construction outlined in Figure \ref{construction}, we construct a QSS$\mathcal{M}$.

The $B=1$ baby Skyrmion has a circularly symmetric energy density. The solution takes the form
\begin{equation} \label{B1}
W_{B=1}(\theta,p,\lambda; z = x +i y) = \frac{ \lambda e^{i \theta} }{z- p} F(|z-p|) \, ,
\end{equation}
where the coordinates $p \in \mathbb{C}$ and $\theta \in S^1$ describe the moduli of the Skyrmion; the position and phase respectively. The position arises from the translational symmetry of the theory while the phase comes from the isorotational symmetry. Our profile function $F$ serves to localize the Skyrmion. It is related to the commonly used profile function $f$ (used in, e.g., \cite{PSZ94} and \cite{WH00}) by
\begin{equation*}
F(r) = r \tan\left(f(r)/2\right) \, .
\end{equation*}
The baby Skyrmions also have an important non-zero mode, known as the breather. This allows the Skyrmion to increase and decrease in size. In actuality, breathing corresponds to a nonlinear change to the profile function, but we can approximate this motion by including $\lambda$ as in \eqref{B1}. The profile function $F$ is fixed so that $\lambda=1$ gives rise to the energy minimizing solution. Overall, the moduli space of the single baby Skyrmion is
\begin{equation*}
\mathcal{M}_1 \cong \mathbb{R}^2 \times S^1 \, .
\end{equation*}

\begin{figure}[h!]
	\begin{center}
		\includegraphics[width=0.8\textwidth]{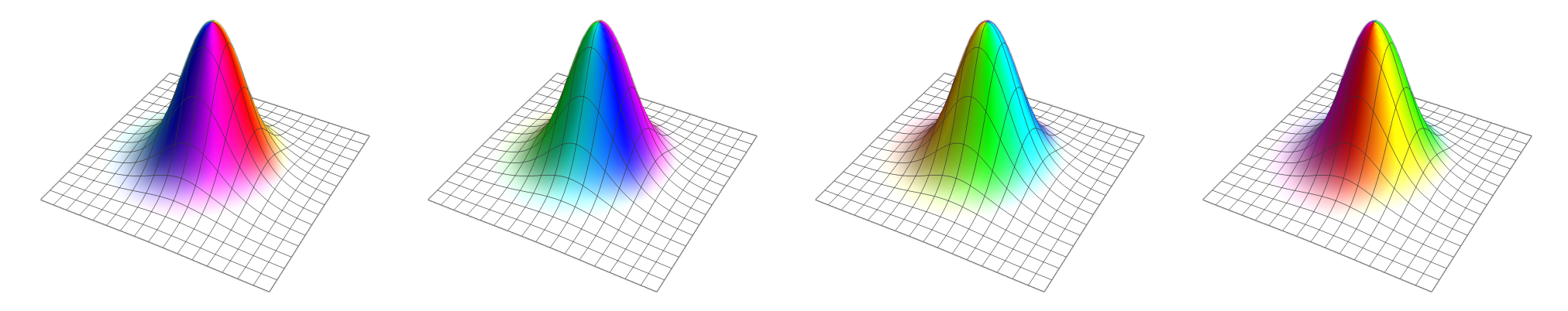} 
	\end{center}
	\caption{ An energy density plot of the circularly symmetric $B=1$ baby Skyrmion, in four different orientations $\theta=0,\pi/2,\pi,3 \pi/2$. Note that the energy density is unaffected by changes in phase.  } \label{B1plot}
\end{figure}

We can graphically represent the phase of the Skyrmion using colors. First, we plot the energy density then color it depending on the value of $W$ at each point. The phase of $W$ is used to give the hue of the color, defined so that $\arg(W) = 0$ is red. This means that $\arg(W) = 2\pi/3$ and $4\pi/3$ correspond to green and blue respectively. We take the magnitude of $W$ to give the opacity, so that the vacuum is white and the center of the Skyrmion is brightly colored. This is a new way to visualize baby Skyrmions though comes from a close analogy with how one visualizes 3-dimensional Skyrmions. We plot the $B=1$ baby Skyrmion for a range of phases in Figure \ref{B1plot}.

A $B=2$ baby Skyrmion configuration can be generated using a superposition approximation, 
\begin{equation} \label{B2}
W_{B=2}(\theta_1, \theta_2, p_1, p_2,\lambda_1, \lambda_2; z) = W_{B=1}(\theta_1,p_1,\lambda_1; z) + W_{B=1}(\theta_2,p_2,\lambda_2; z) \, .
\end{equation}
This gives an approximate static solution when $|p_1 - p_2|$ is large and $\lambda_1=\lambda_2=1$. Although the Skyrmion solutions have a preferred size, it will be instructive to allow $\lambda_1$ and $\lambda_2$ to vary. 

\begin{figure}[h!]
	\begin{center}
		\includegraphics[width=0.9\textwidth]{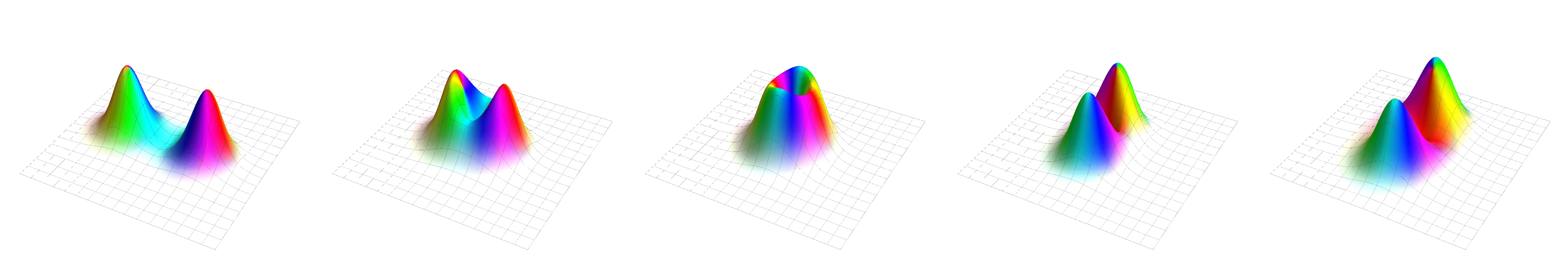} 
	\end{center}
	\caption{ Scattering in the attractive channel. We plot the energy density of the baby Skyrmions at several time steps and time evolution is read left to right. The Skyrmions start at rest, before moving towards each other. They pass through a circularly symmetric configuration and depart on a path perpendicular to the initial path, before finally coming to rest.} \label{AtSca}
\end{figure}

\begin{figure}[h!]
	\begin{center}
		\includegraphics[width=0.9\textwidth]{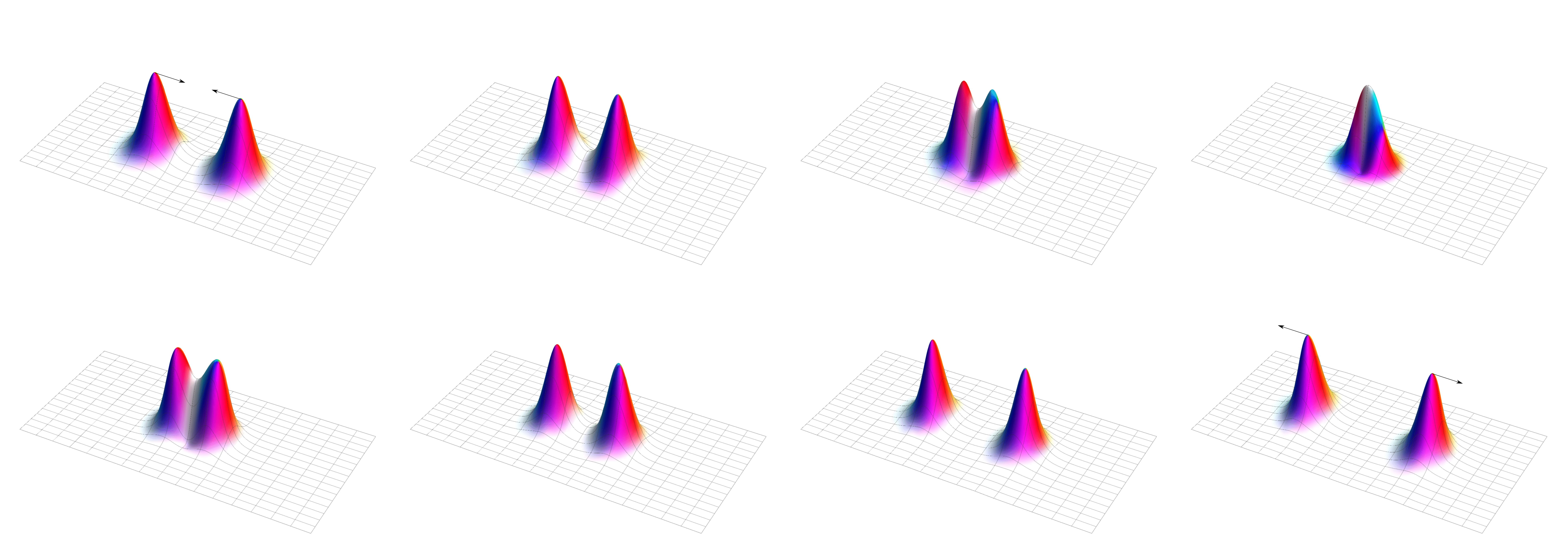} 
	\end{center}
	\caption{ Scattering in the unattractive channel. The Skyrmions are boosted towards each other (as indicated by the arrows in the first sub-figure), and deform when close. After forming a compact high-energy object, they bounce back, moving away from each other until reaching infinity (the side of our box). This demonstrates that the unattractive channel is unattractive.  } \label{UnSca}
\end{figure}

Before attempting to construct the two-Skyrmion QSS$\mathcal{M}$, we shall get some intuition for the classical scatterings. The asymptotic interaction potential between two baby Skyrmions is
\begin{equation*}
V_{\text{int}} \propto \cos(\theta_1 - \theta_2) \, .
\end{equation*}
The Skyrmions attract most strongly when $\theta_1 - \theta_2 = \pi$. This is known as the attractive channel. Here, the point of closest contact between the Skyrmions has the same pion field structure (or coloring, graphically). A scattering using this initial data is shown in Figure \ref{AtSca}. As the Skyrmions approach, they deform until obtaining circular symmetry, before scattering out at right angles. This $90^o$ scattering is well known in a variety of soliton models. The unattractive channel, where $\theta_1 - \theta_2 =0$, is less studied. If the initial Skyrmions have no velocity, they are repelled by each other and race away to infinity. For large initial velocities, they approach each other and deform but never pass through each other. One can show this by tracking the positions of the Skyrmions (that is, where $\phi_3 = -1$). For the paths we numerically generate, such as the one shown in Figure \ref{UnSca}, the Skyrmions never leave the $x$-axis. 

To discuss the two-Skyrmion QSS$\mathcal{M}$ in detail, we should first examine the structure of $\mathcal{M}^\text{asy}_{1,1}$. This contains the configurations approximately given by $W_{B=2}(\theta_1, \theta_2, p_1, p_2,1,1; z)$ when $|p_1 - p_2|$ is large. Many of our results depend on a surprising fact: every configuration in $\mathcal{M}^\text{asy}_{1,1}$ has a reflection symmetry. This was pointed out by Piette, Schroers and Zakrzewski in \cite{PSZ94} but the consequences were not fully explored. To show the existence of the symmetry, we need to show that a physical reflection can be compensated by an iso-reflection. This is most easily seen using the map \eqref{B2}. The symmetry is a reflection in the line perpendicular to the line joining the Skyrmion centers. This maps the original configuration to a configuration of two antiSkyrmions (where $B$ is negative). In terms of colors, the order of the colors which appear on the Skyrmions has reversed. This mapping on the color (target) space has two fixed points, separated by $\pi$. An isoreflection around a line joining these points in color  space returns the configuration to its initial state. In terms of the coordinates on $\mathcal{M}_{1,1}^\text{asy}$, this two step process is
\begin{equation*}
\underbrace{z \to p_2 + \frac{p_1 - p_2}{\bar{p}_1 - \bar{p}_2}\left(\bar{p}_1 - \bar{z}\right)}_\text{physical relfection} \, \text{ and } \underbrace{ W \to -e^{i(\theta_1 + \theta_2)}\frac{\bar{p}_1 - \bar{p}_2}{p_1-p_2} \, \overline{W} }_\text{iso-reflection} \, .
\end{equation*}
and so the statement of reflection symmetry is 
\begin{align} \label{refsymm}
W_{B=2}(\theta_1,&\theta_2,p_1,p_2,1,1;z) = \\ &-e^{i(\theta_1 + \theta_2)}\frac{\bar{p}_1 - \bar{p}_2}{p_1-p_2} \, \overline{W}_{B=2}\left(\theta_1,\theta_2,p_1,p_2,1,1; p_2 + \frac{p_1 - p_2}{\bar{p}_1 - \bar{p}_2}\left(\bar{p}_1 - \bar{z} \right) \right) \, . \nonumber
\end{align}

We now factor out the moduli contained in $\mathcal{M}^\text{asy}_{1,1}$, explicitly introducing coordinates to do so. We may use translational symmetry to set the center of mass to be zero by insisting that $p_2 = -p_1 = p$, rotational symmetry to place the Skyrmions on the $x$-axis, meaning that $p\in \mathbb{R}$, and  isorotational symmetry to set $\theta_2 = -\theta_1 = \theta$. Maps of this kind take the form
\begin{equation*}
W_\text{B=2}(\theta,p; z) = \frac{e^{i \theta} }{z - p}F(|z-p|) + \frac{e^{-i\theta}}{z+p}F(|z+p|) \, .
\end{equation*}  
These configurations parametrize the two-dimensional reduced asymptotic submanifold $\widetilde{\mathcal{M}}_{1,1}^\text{asy}$ . The  coordinate $p\in \mathbb{R}$ represents the relative position and $\theta$ the relative phase of the Skyrmions. These interpretations are valid when the Skyrmions are far apart. Note that $\theta=0$ and $\theta=\pi / 2$ correspond to the unattractive and attractive channels respectively. Configurations with different values of $\theta$ and $p$ have will have different energies. Since the potential is not flat in these directions (unlike in the direction of the moduli) these coordinates are often called massive. Overall, the asymptotic submanifold is six-dimensional and takes the form
\begin{equation} \label{Masyprod}
\mathcal{M}^\text{asy}_{1,1} = G\times \widetilde{\mathcal{M}}^\text{asy}_{1,1} \cong\underbrace{E_2 \times S^1}_\text{moduli} \times \underbrace{S^1 \times \mathbb{R}^+}_\text{massive modes} \, .
\end{equation}

We can now consider the time evolution of configurations from $\widetilde{\mathcal{M}}^\text{asy}_{1,1}$. To do so, it is helpful to consider a larger space of configurations; where the individual Skyrmions can alter their sizes. Factoring out the moduli as above, there is a four-dimensional manifold of configurations, parametrized by
\begin{equation*}
W_{B=2}^\text{in} = \lambda_1 e^{i\theta}\frac{1}{z-p}F(|z-p|) + \lambda_2 e^{-i\theta}\frac{1}{z+p}F(|z+p|) \, .
\end{equation*}
These are our initial field configurations. The initial field momentum is taken so that the Skyrmions move towards one another at velocity $v$. This amounts to taking the position to be linearly dependent on time, $p = p_0 - v t$. The initial configurations are on the $x$-axis and so the reflection symmetry \eqref{refsymm} is realized as 
\begin{equation} \label{ref2}
W^\text{in}_{B=2}(\theta,p,\lambda_1,\lambda_2;z) = -\overline{W}^\text{in}_{B=2}(\theta,p,\lambda_1,\lambda_2;-\bar{z})  \, .
\end{equation}
In terms of the map, noting that $p$ is real, we find
\begin{align*}
W^\text{in}_{B=2}(z) &= \lambda_1 e^{i\theta}\frac{1}{z-p}F(|z-p|) + \lambda_2 e^{-i\theta}\frac{1}{z+p}F(|z+p|)  \\
&= \lambda_2 e^{i\theta}\frac{1}{z-p}F(|z-p|) + \lambda_1 e^{-i\theta}\frac{1}{z+p}F(|z+p|) = -\overline{W}^\text{in}_{B=2}(-\overline{z}) \, ,
\end{align*}
which implies that $\lambda_1 = \lambda_2$. Physically, the reflection symmetry demands that the Skyrmions  have the same size.

The symmetry is preserved in time evolution and so has dynamical consequences. We used rotational symmetry to initially place the Skyrmions on the $x$-axis and so the reflection symmetry is reflection in the $y$-axis. As such, if the Skyrmions scatter and separate after the collision, they must either scatter at right angles (as in Figure \ref{AtSca}) or bounce back (as in Figure \ref{UnSca}). Assuming the Skyrmions do scatter after collision, we can see what the reflection symmetry implies. Noting that $p$ is now imaginary (since the Skyrmions come out on the $y$-axis) and that we cannot guarantee the phases are related in the same way as they were initially, we find
\begin{align*}
W^\text{out}_{B=2}(z) &= \lambda_1 e^{i\theta_1}\frac{1}{z-p}F(|z-p|) + \lambda_2 e^{-i\theta_2}\frac{1}{z+p}F(|z+p|)  \\
&= \lambda_1 e^{-i\theta_1}\frac{1}{z-p}F(|z-p|) + \lambda_2 e^{i\theta_2}\frac{1}{z+p}F(|z+p|) = -\overline{W}^\text{out}_{B=2}(-\overline{z}) \, .
\end{align*}
Now there is no restriction on the sizes. Instead, $\theta_1 - \theta_2 = n \pi$ and the Skyrmions have fixed relative phase: they are either in the attractive or unattractive channel. Hence the superposition approximation predicts that Skyrmions with the same size colliding head-on, out of the attractive channel, will scatter at right angles or bounce back. If they scatter at right angles, the outgoing Skyrmions will then be fixed in the (un)attractive channel though can have different sizes. 

\begin{figure}[h!]
	\begin{center}
		\includegraphics[width=1.0\textwidth]{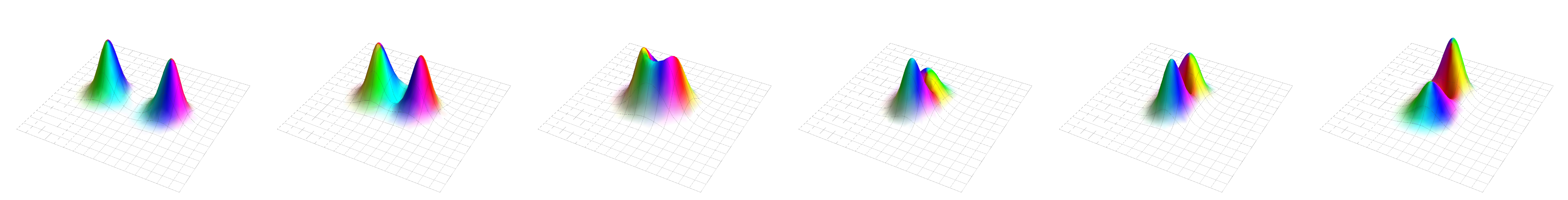} 
	\end{center}
	\caption{ Scattering in between the attractive and unattractive channels. The Skyrmions are boosted towards each other (with velocity $v=0.05$). They scatter through a deformed ring structure before emerging at right angles, one larger than the other. The mode corresponding to relative size fluctuations is excited and the smaller Skyrmion grows with respect to the other. Note that the outgoing Skyrmions are in the attractive channel, the color red showing at both points of closest contact.} \label{InSca}
\end{figure}

To check this prediction, we numerically simulated several scatterings using full field dynamics. We used many different initial conditions and a typical simulation is shown in Figure \ref{InSca}. The dynamics were generated using methods similar to those detailed in \cite{lotsanums}. One can clearly see that the prediction from the superposition approximation is confirmed: the outgoing Skyrmions have different sizes. This is a new physical mechanism - it converts potential energy from a relative orientation mode to a relative breathing mode. When the solitons are widely separated, the relative orientation is a zero mode while the relative size is a massive mode. That these motions mix is unexpected, and softens the dichotomy between zero and non-zero modes. If applied to Skyrmions in nuclear physics, it shows a way to easily generate a roper resonance (modeled as a breathing excitation \cite{HH84,AHRW18}) from nucleon-nucleon scatterings.

Having understood the dynamics, we can now prove the following: a 2 baby Skyrmion QSS$\mathcal{M}$ must have dimension greater than six. Let us take $\mathcal{S}_{1,1}^\text{in} \cong \mathcal{M}^\text{asy}_{1,1}$, the minimal ingoing manifold. This has dimension six. We have just argued, using the superposition approximation and numerical simulations, that the outgoing configuration space $\mathcal{T}_{\infty}[\mathcal{S}_{1,1}^\text{in}]$ does not contain a relative phase degree of freedom - the outgoing configurations are fixed in the (un)attractive channel. Hence it does not contain $\mathcal{M}^\text{asy}_{1,1}$, and we must add additional configurations to the manifold. We cannot remove any configurations from $\mathcal{S}_{1,1}^\text{in}$, since it contains the minimal set of ingoing configurations. Consequentially its dimension must increase. 

Moreover, we understand the outgoing configurations which come from the reduced asymptotic submanifold, $\mathcal{T}_{\infty}[\widetilde{\mathcal{M}}^\text{asy}_{1,1}]$. Ingoing configurations with a relative phase difference get mapped to outgoing configurations with a relative size difference. Hence the manifold of outgoing configurations is 
\begin{equation}
\mathcal{T}_{\infty}[\widetilde{\mathcal{M}}^\text{asy}_{1,1}] \cong  \mathbb{R}^+ \times \mathbb{R}^+\, ,
\end{equation}
where the factors are interpreted as the relative size and relative position of the outgoing Skyrmions. Hence the combination of this with $\widetilde{\mathcal{M}}^\text{asy}_{1,1}$ gives a consistent ingoing three-dimensional manifold
\begin{equation} \widetilde{\mathcal{S}}_\text{in} \cong \mathbb{R}^+ \times \mathbb{R}^+ \times S^1\, ,
\end{equation}
where these factors represent the relative size, relative position and relative phase. Once we add the four-dimensional fiber, representing the moduli, the total space is seven-dimensional. Hence the smallest 2-baby Skyrmion manifold has at least seven dimensions. The seven-dimensional ingoing and outgoing manifolds can be parametrized by
\begin{equation} \label{allB2}
W_{B=2} = \lambda e^{i\theta_1}\frac{1}{z-p_1}F(|z-p_1|) + \lambda^{-1} e^{i\theta_2}\frac{1}{z-p_2}F(|z-p_2|) \, ,
\end{equation}
The previously missing configurations in the outgoing manifold (with non-zero relative phase) are generated by ingoing configurations with non-zero relative size.

Note that the phase parameter takes values in $S^1$ while the size parameter takes values in $\mathbb{R}^+$. We claim that these parameters map onto one another before and after scattering, but the sets are not isomorphic. However, we claimed earlier that Skyrmions in the unattractive channel always bounce back. This means that this ingoing configuration does not map to an outgoing configuration. Hence the map is really between $S^1\setminus\{\boldsymbol{0}\} \cong \mathbb{R}^+$. To prove this one must show that Skyrmions in the unattractive channel cannot scattering at right angles. In the attractive channel, Skyrmions pass through the ``central" circularly symmetric configuration. One might be able prove that no such ``central" configuration exists in the unattractive channel, due to the symmetries present. We leave this for future research. The superposition approximation told us that the outgoing Skyrmions were in either the attractive or unattractive channels. Our numerical simulations suggest that they always emerge in the attractive channel. This can probably be proved using the continuity of configuration space. 

Having understood the structure of the 2-baby Skyrmion configuration space, it would be interesting to construct the space explicitly, perhaps using \eqref{allB2} as a starting point. The space has previously been modeled using degree two Rational Maps (RMs) \cite{Sut91}, successful when the second term in the Lagrangian \eqref{Lag} is tuned to be small. The link between the superposition and RM approximations is clear if we rewrite \eqref{allB2} as
\begin{equation*}
W_{B=2} = \frac{ z \left(\lambda e^{i\theta_1} F_1 + \lambda^{-1}e^{i \theta_2} F_2 \right) - \left(\lambda e^{i\theta_1}p_2 F_1 + \lambda^{-1} e^{i \theta_2} p_2 F_2\right)}{(z-p_1)(z-p_2)} \, ,
\end{equation*}
where $F_i = F(|z-p_i|)$. This almost takes the form of a degree two Rational Map
\begin{equation*}
W_{B=2}(z) = \frac{a z + b}{(z-c)(z-d)} \, ,
\end{equation*} 
where $a,b,c,d \in \mathbb{C}$. The issue is that $F_1$ and $F_2$ are not holomorphic. They provide the localization of the Skyrmion and the RM approximation must be modified to account for this. If this is done, baby Skyrmion-Skyrmion scattering can be studied in detail. Many questions arise -- is the space geometrically complete (in contrast to the 2-lump space)? Is it chaotic? How does the transfer of ``relative phase energy" to ``size deformation energy" occur in practice? One can then try to solve the quantum scattering problem. No non-BPS quantum soliton scattering problem has been completed so this would be the first. It would provide essential intuition for the physically important Skyrmion-Skyrmion scattering, which describes low energy nucleon-nucleon scattering.

Before moving on, let us summarize the calculation. We studied the time evolution of $\widetilde{\mathcal{M}}^\text{asy}_{1,1}$ for two baby Skyrmions. Using the superposition approximation to construct an ingoing manifold, we were able to predict the set of outgoing configurations evolved from the ingoing set. Numerical calculations confirmed this prediction. This also revealed what else to include in the ingoing manifold: relative size fluctuations. These included, we formed a consistent seven-dimensional approximation for the QSS$\mathcal{M}$, displayed in equation \eqref{allB2}. The next step would be to explicitly generate the manifold.

\section{The Skyrme model}

This Section closely follows Section 4. We introduce the Skyrme model then discuss the properties of a 2-Skyrmion QSS$\mathcal{M}$.

In the Skyrme model, proposed  by Tony Skyrme in the early 60s, nuclei are described as solitons (Skyrmions) in a nonlinear theory of pions \cite{Sk61}. Its Lagrangian is
 \begin{equation} \label{LagS}
 \mathcal{L} =  \partial_\mu\boldsymbol{\pi} \cdot \partial^\mu \boldsymbol{\pi} - \tfrac{1}{2}\left( \left(\partial_\mu \boldsymbol{\pi} \cdot \partial^\mu \boldsymbol{\pi}\right)^2 - \left(\partial_\mu \boldsymbol{\pi} \cdot \partial^\nu \boldsymbol{\pi} \right)\left(\partial_\nu \boldsymbol{\pi} \cdot \partial^\mu \boldsymbol{\pi}\right) \right) - m_\pi^2(1-\pi_4) \, ,
 \end{equation}
where $m_\pi$ is the dimensionless pion mass, $\pi_i$ are the pion fields for $i=1,2,3$ and $\pi_4$ is an auxiliary field which enforces the constraint $\boldsymbol{\pi}\cdot\boldsymbol{\pi} = 1$. The Lagrangian enjoys Lorentz and isorotational ($\pi_i \to A_{ij}\pi_j, \, A \in SO(3)$) symmetries. Hence, the fiber of our quantum scattering space will be $E_3 \times SO(3)$. The analogy with the baby Skyrme model is apparent at the level of the Lagrangian. Now $\boldsymbol{\pi}$ is a map from $\mathbb{R}^3 \to S^3$ but the finite energy condition causes a one-point compactification of space. Hence finite energy configurations are maps between three-spheres (rather than two-spheres, which appear in the Baby Skyrme model) and these maps have a conserved integer $B$. A $B$-Skyrmion is the energy minimizing solution with charge $B$.

The $B=1$ Skyrmion has a spherical energy density. The solution takes the form
\begin{equation*}
\left(\pi_i(X,A;\boldsymbol{x}), \pi_4(X,A;\boldsymbol{x})\right) = \left(A_{ij}\widehat{(x-X)}_j \cos\left( g(|x-X|) \right) , \sin\left( g(|x-X|) \right) \right) \, .
\end{equation*}
The coordinate $X \in \mathbb{R}^3$ describes the position of the Skyrmion while the iso-rotational matrix, $A \in SO(3)$, determines its orientation. The function $g$ is known as the profile function and satisfies an ODE determined by the equations of motion derived from \eqref{LagS}. The one-Skyrmion moduli space is then
\begin{equation*}
\mathcal{M}_1 \cong \mathbb{R}^3 \times \text{SO}(3) \, ,
\end{equation*}
and has dimension six.

\begin{figure}[h!]
	\begin{center}
		\includegraphics[width=1.0\textwidth]{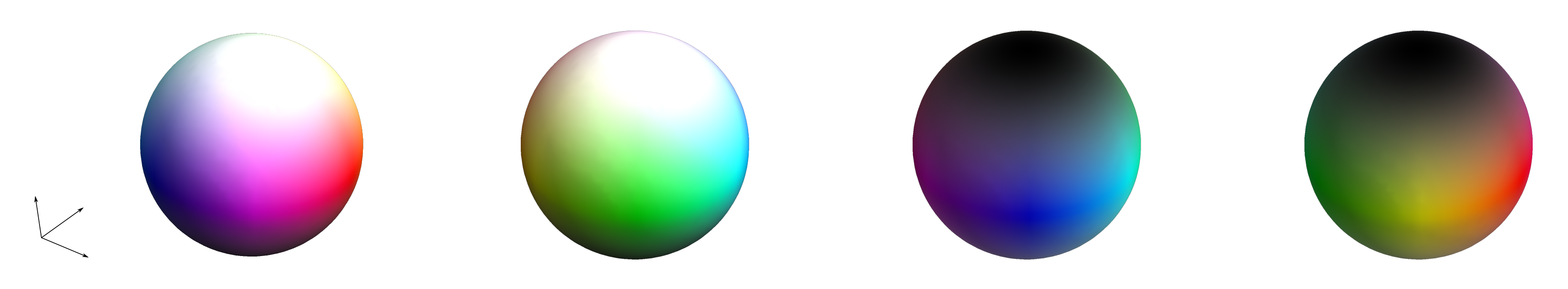} 
	\end{center}
	\caption{The $B=1$ Skyrmion in several orientations. The configurations from left to right correspond to the isorotational matrix $A$ being diagonal with non-zero entries $(1,1,1)$, $(-1,-1,1)$, $(1,-1,-1)$ and $(-1,1,-1)$.  } \label{B1s}
\end{figure}

A Skyrmion can be visualized by plotting a contour of its energy density then coloring it depending on the value of $\boldsymbol{\pi}$ on the contour. It is colored using the Runge color sphere. The Skyrmion is white/black when $\pi_3$ equals ±1 and red, green and blue when $\pi_1 + i\pi_2$ is equal to 1, $\exp(2\pi i/3)$ and $\exp(4\pi i/3)$ respectively. A Skyrmion is plotted in several orientations in Figure \ref{B1s}. This coloring was originally proposed in \cite{FLM13}.

We can generate a $B=2$ Skyrme configuration using the product approximation. This is most easily expressed by first writing the Skyrme field as an $SU(2)$-valued field $U$, 
\begin{equation}
U(X,A;\boldsymbol{x}) =  \pi_4(X,A;\boldsymbol{x}) \boldsymbol{1}_2 + i \sum_i \tau_i \pi_i(X,A;\boldsymbol{x}) \, ,
\end{equation}
where $\tau_i$ are the Pauli matrices and $\boldsymbol{1}_2$ is the $2\times2$ identity matrix. The product approximation is then
\begin{equation} \label{prodaz}
U_{B=2}(X_1,X_2,A_1,A_2; \boldsymbol{x} ) = U_{B=1}(X_1,A_1; \boldsymbol{x} )  U_{B=1}(X_2,A_2; \boldsymbol{x} )  \, .
\end{equation}
This gives a good approximation to solutions of the field equations when $|X_1 - X_2| >> 1$ but breaks down when the Skyrmions approach one another. There is also the symmetrized product approximation
\begin{align} \label{prodsymm}
U_{B=2}(X_1,X_2,A_1,A_2) =  \Big(&U_{B=1}(X_1,A_1 )  U_{B=1}(X_2,A_2) \nonumber
 \\&+ U_{B=1}(X_2,A_2 )  U_{B=1}(X_1,A_1 ) \Big) N^{-1} \, ,
\end{align}
where $N$ normalizes the field \cite{NR88}. This has the advantage of being symmetric under interchange of particles but $N$ can be non-zero when the Skyrmions are brought close together, creating a singularity in the field. The configurations in the 12-dimensional asymptotic submanifold can be approximately generated using \eqref{prodaz} or \eqref{prodsymm}. It's structure is
\begin{equation*} 
\mathcal{M}_{1,1}^\text{asy} \cong (\mathcal{M}_1 \times \mathcal{M}_1)\cong ( \mathbb{R}^3 \times SO(3) \times \mathbb{R}^3 \times SO(3) ) \, .
\end{equation*}
Using either approximation we can calculate an asymptotic interaction potential. Once again, there is an energetically preferred set of configurations called the attractive channel. A pair of Skyrmions are in the attractive channel if they are colored the same at their point of closest contact. When  Skyrmions relax numerically in the attractive channel (via gradient flow, for example), they form the $B=2$ Skyrmion solution. This has toroidal symmetry and is not described by either product approximation.

Our goal is to find a QSS$\mathcal{M}$ for the 2-Skyrmion system. This is essential for describing nucleon-nucleon scattering in the Skyrme model and could also be important for nucleon-nucleon interactions. Past work in this area has generally taken the product approximation and applied it at all separations \cite{JJ85}. The approximation generates a 12-dimensional configurations space, which contains a copy of $\mathcal{M}^\text{asy}_{1,1}$. However, we know that this approximation does not include the toroidal configuration, which is the minimum energy configuration in the $B=2$ sector. Atiyah and Manton suggested another approach where the Skyrmion configuration space is modeled by the instanton moduli space \cite{AM93}. This method includes the toroidal $B=2$ Skyrmion solution as well as the asymptotic configurations \eqref{prodaz}. However, there is choice in what the total configuration space should be. There is a $16$-dimensional space of instantons, though the generation of the Skyrmion ``uses up" one of these, so there  is only a $15$-dimensional space of instanton-generated Skyrmions. Atiyah and Manton found a $12$-dimensional submanifold, $\mathcal{M}_{12}$, of this $15$-dimensional space. This manifold also arises as the union of gradient flow paths from the unstable $B=2$ spherically symmetric Skyrmion \cite{Man88}. We will now show that $\mathcal{M}_{12}$ cannot be a QSS$\mathcal{M}$ and so is not appropriate for describing nucleon-nucleon scattering. It could still be a useful tool for investigating the nucleon-nucleon force.

\begin{figure}[h!]
	\begin{center}
		\includegraphics[width=1.0\textwidth]{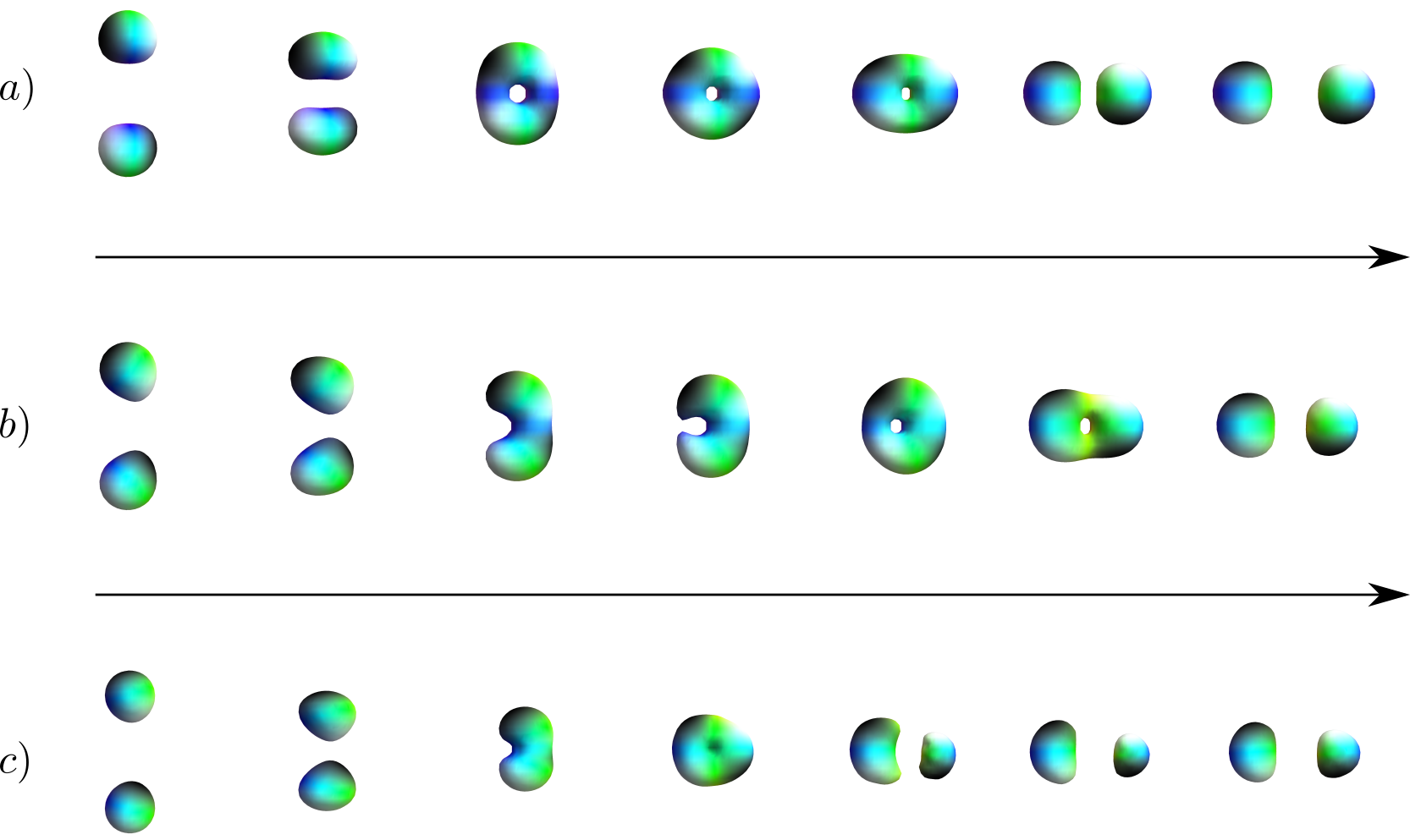} 
	\end{center}
	\caption{ Three scatterings of two Skyrmions. Scattering $a)$ occurs in the attractive channel and hence the outgoing Skyrmions have the same size. Scatterings $b)$ and $c)$ begin out of the attractive channel. The latter is very close to the unattractive channel and as such the outgoing Skyrmions have a large size difference. } \label{Scatterings}
\end{figure}

Suppose we describe the ingoing part of a two-Skyrmion QSS$\mathcal{M}$, denoted $\mathcal{S}_{1,1}$, as a 12-dimensional manifold. This is the same size as the asymptotic submanifold $\mathcal{M}_{1,1}^\text{asy}$ and so
\begin{equation*}
\mathcal{S}_{1,1}^\text{in} \cong \mathcal{M}_{1,1}^\text{asy} \, .
\end{equation*}
For this to be a QSS$\mathcal{M}$, the outgoing configurations must contain all configurations from $\mathcal{M}_{1,1}^\text{asy}$. We can factor out the moduli in $\mathcal{S}^\text{in}_{1,1}$: translations, rotations and isorotations. This leaves a three-dimensional non-trivial manifold of configurations not related by moduli, which all have different energies. Physically, the three degrees of freedom are relative orientations (two) and separation (one). Note that you would naively expect three orientation degree of freedom, but one of these is equivalent to a rotation. There is then a two-dimensional geodesic submanifold with a reflection, analogous to the reflection symmetry present for baby Skyrmions. These are the configurations where the Skyrmion's orientations match at least once. Three such configurations are shown in the leftmost column of Figure \ref{Scatterings}.  Here, the matching color is teal and it occurs on the point of the Skyrmion facing the reader. Using the intuition gained from the baby Skyrmion model we can make a prediction: Skyrmions whose initial data have this reflection symmetry will scatter at $90^o$, coming out in the attractive channel with different sizes.

To test this prediction, we simulated head-on collisions  of Skyrmions using full field dynamics. The interaction energy on $\mathcal{C}_2^\infty$ encourages the Skyrmions to dynamically move into the attractive channel. If the Skyrmions are let go at rest, they will often reorient before the collision occurs. To overcome this difficulty, we boost the initial configurations at each other. Three simulations are shown in Figure \ref{Scatterings}, which confirm the prediction.

With the dynamical prediction confirmed, we are done. The outgoing configurations $\mathcal{T}_\infty(\mathcal{S}_{1,1}^\text{in})$ contain a degree of freedom which describes the relative size of the Skyrmions. The inclusion of this degree of freedom must come at the exclusion of another. Hence $\mathcal{M}^\text{asy}_{1,1} \nsubseteq \mathcal{T}_\infty(\mathcal{S}_{1,1}^\text{in})$ and there is no 12-dimensional QSS$\mathcal{M}$. An appropriate manifold should also include relative size fluctuations of the Skyrmions. This means that a 2-Skyrmion QSS$\mathcal{M}$ is at least 13-dimensional. Including size fluctuations for both Skyrmions would give a 14-dimensional manifold of configurations. This would agree with recent numerical work by the author and Gudnason \cite{HG18}. Here, small fluctuations around the $B=2$ torus were studied and exactly 14 modes were found.

Now that we know how large $\mathcal{S}_{1,1}$ should be, we can consider how to construct a 14-dimensional QSS$\mathcal{M}$ explicitly. The instanton approximation might be flexible enough to describe it. The ingoing asymptotic manifold is simple to write down within the approximation. However, it is not known if the instanton approximation accurately describes Skyrmion motion out of the attractive channel. In fact, very little is known about this motion, even numerically. From our work, we now know that Skyrmions of the type shown in Figure \ref{Scatterings} change size after collision, but this is only a two-dimensional geodesic submanifold of the 3-dimensional non-trivial manifold of configurations. We had nothing to say about other scatterings. These should be studied and compared to the instanton predictions. If the comparison was favorable, the instanton-generated Skyrmions could then be used to study nucleon-nucleon scattering in the Skyrme model.

Another, somewhat naive, idea to describe the manifold comes from an analogy with lumps. The (centered) two lump solutions can be described using a degree two Rational Map with a linear denominator. This can be written as a sum of poles,
\begin{equation*}
W(z) = \frac{a z+ b}{z^2 - c^2} = \frac{b+a c}{z-c}\frac{1}{z-c} - \frac{b-ac}{2c}\frac{1}{z+c} = \frac{q_1}{z-p_1} + \frac{q_2}{z-p_2} \, ,
\end{equation*} 
so that the two-lump solution can be thought of as a pair of positions $p_i$ and internal coordinate $q_i$ (which describe sizes and phases). Similarly, an $n$-lump can be written as a sum of $n$ poles with positions $p_i$ and residues $q_i$. 
We posit a similar structure for Skyrmions. A Skyrmion has position $X_i \in \mathbb{R}^3$ and an internal coordinate given by a quaternion $Q_i$. The magnitude of $Q_i$ should describe the size of the Skyrmion with $\hat{Q}_i$ giving its SU(2) orientation. One then needs a mapping from the set $\{X_i,Q_i\}$ to Skyrme configurations.

\section{Conclusions and Further Work}

In this paper, we have proposed a definition and a construction of a quantum soliton scattering manifold (QSS$\mathcal{M}$); the minimal manifold required to describe quantum multisoliton scattering. The construction is simple: one takes the union of all classical scattering paths which take their initial data from the asymptotic submanifold $\mathcal{M}^\text{asy}_{1,1}$. If the late-time configurations contain new configurations (not seen in $\mathcal{M}^\text{asy}_{1,1}$), a mistake has been made and these should have been included from the start. We applied this idea to two- and three-dimensional Skyrmions, showing that the late-time configurations include a relative size orientation, not seen in $\mathcal{M}^\text{asy}_{1,1}$. A QSS$\mathcal{M}$ should contain these degrees of freedom and so its dimension is larger than one expect from BPS intuition. Although the construction appears to depend on a detailed knowledge of the classical paths, our work has shown that you can make progress without this. In the Skyrme model we only considered paths on a geodesic submanifold of the whole configurations space and, using these, argued that the $2$-Skyrmion QSS$\mathcal{M}$ is at least $13$-dimensional and have hinted at its global structure. This is an important step in solving the quantum two-body problem in the Skyrme model, which has been discussed and unsolved since Skyrme's original work \cite{Sk61}. The obvious next step is to try and construct the manifold using the instanton approximation or an entirely new idea.

The method is rather general and we'll now describe some further applications, first for larger Skyrmions. Suppose we want to describe $\alpha-\alpha$ scattering in the Skyrme model. This is modeled by the scattering of two $B=4$ Skyrmions. Since the particle content we're interested in are $\alpha$-particles, the asymptotic submanifold $\mathcal{M}_{4,4}^\text{asy}$ is the product of the moduli spaces of the $B=4$ Skyrmions. This is well understood and can be constructed using the product approximation. We can then explore all scattering paths whose initial data are in $\mathcal{M}_{4,4}^\text{asy}$. If the late-time configurations are not contained in $\mathcal{M}_{4,4}^\text{asy}$, the additional degrees of freedom should also be included. With luck, one has access to an approximation scheme, such as the instanton approximation, and can construct the manifold numerically. Once done, the problem is reduced to quantum mechanics on the manifold.

These methods may also apply to other physical systems. A possible example is black hole dynamics\footnote{I thank Peter Forg\'acs for the suggestion}. Colliding black holes which do not rotate and have maximal charge can be described by motion on a moduli space. To get close to physically interesting black holes one must slacken these constraints and hence slacken the moduli space picture. Our method might reveal how to do so consistently - perhaps by including additional degrees of freedom.

\section*{Acknowledgments}

 I would like to thank Bjarke Gudnason, Jonathon Rawlinson, Martin Speight, Thomas Winyard and especially Derek Harland for useful discussions throughout the development of this project. The author is supported by The Leverhulme Trust as an Early Careers Fellow.

\end{document}